\documentclass[12pt]{iopart}

\usepackage{hyperref}
\usepackage{wasysym}
\usepackage{bm}

\usepackage{graphicx,color}

\definecolor{mygreen}{rgb}{0,0.5,0} 
\definecolor{myblue}{rgb}{0,0,0.75} 
\definecolor{mymagenta}{cmyk}{0,1,0,0.12} 
\definecolor{mycyan}{cmyk}{1,0,0,0.12} 
\newcommand{\mtext}[1]{{\color{myblue}#1}}
\renewcommand{\mtext}[1]{{#1}}
\newcommand{\btext}[1]{{\color{myblue}#1}}
\renewcommand{\btext}[1]{{#1}}

\newcommand{\qtext}[1]{{\color{mymagenta}#1}}
\renewcommand{\qtext}[1]{{\color{mymagenta}}}

\newcommand{\be}{\begin{equation}}
\newcommand{\ee}{\end{equation}}
\newcommand{\bea}{\begin{eqnarray}}
\newcommand{\eea}{\end{eqnarray}}

\newcommand{\ket}[1]{\left|#1\right>}
\newcommand{\bra}[1]{\left<#1\right|}

\newcommand{\expect}[1]{\left<#1\right>}

\newcommand{\nn}{\nonumber \\ }

\newcommand{\nnpropto}{\nonumber \\ & \propto &}
\newcommand{\nnp}{\nonumber \\ & & +}

\newcommand{\nnt}{\nonumber \\ & & \times}

\newcommand{\var}{{\rm var}}

\newcommand{\PRLsection}[1]{\noindent {\it#1} -}

\newcommand{\SM}{S{\o}rensen-M{\o}lmer}

\newcommand{\nc}{n_{\rm c}}
\renewcommand{\ns}{n_{\rm s}}
\newcommand{\hermsym}{\cdot}

\renewcommand{\PRLsection}{\section}

\newcommand{\NGME}{10}

\begin{document}

\title{Extreme spin squeezing for photons}
\author{Morgan W.~Mitchell$^{1,2}$, Federica A. Beduini$^1$}
\address{$^1$ICFO-Institut de Ciencies Fotoniques, Mediterranean Technology Park, 
08860 Castelldefels (Barcelona), Spain} 
\address{$^2$ ICREA-Instituci\'{o} Catalana de Recerca i Estudis Avan\c{c}ats, Lluis Companys 23, 08010 Barcelona, Spain}
\ead{morgan.mitchell@icfo.es}
\begin{abstract} 
We apply spin-squeezing techniques to identify and quantify highly multi-partite photonic entanglement in polarization-squeezed light. We consider a practical single-mode scenario, and find that Wineland-criterion polarization squeezing implies entanglement of a macroscopic fraction of the total photons.   A Glauber-theory computation of the observable $N$-photon density matrix, with $N$ up to 100, finds that $N$-partite entanglement is observable despite losses and without post-selection.  Genuine multi-partite entanglement up to at least $N = \NGME$ is similarly confirmed.  The preparation method can be made intrinsically permutation-invariant, allowing highly efficient state reconstruction.  In this scenario, generation plus detection requires $O(N^{0})$ experimental resources, in stark contrast to the typical exponential scaling. We estimate existing detectors could observe 1000-partite entanglement from a few dB of polarization squeezing.  
\end{abstract} 
\submitto{\NJP} 
\maketitle

\section{Introduction}

{
The theorems known as spin-squeezing inequalities (SSIs) establish a relationship between entanglement and squeezing, providing a bridge between a microscopic and a macroscopic quantum phenomenon, respectively.
An atomic spin ensemble, i.e. a macroscopic collection of identical atoms, can be described both by a micro-state that specifies the correlated behavior of all of its components, and by collective spin operators describing the ensemble as a whole.  While the micro-state is the object of fundamental interest for quantum information and simulation tasks, it is inaccessible to measurement in all but the smallest ensembles.  In contrast, the collective spin operators  can often be measured with high precision \cite{GrossN2010,ZhangPRL2012}.  In some cases, precise non-destructive measurement \cite{SewellNP2013} can also be used to create squeezing \cite{AppelPNAS2009,KoschorreckPRL2010a,KoschorreckPRL2010b,SewellPRL2012}.  

In this experimental context, SSIs provide a unique window into the entanglement properties of spin ensembles.  As a first result, S\o{}rensen et al. \cite{SorensenN2001} showed that for a spin-1/2 ensemble, a collective spin   variance below the standard quantum limit, i.e. spin squeezing, implies pairwise entanglement of at least one pair of spins in the ensemble.  Subsequent SSIs extend this to higher-spin atoms \cite{VitaglianoPRL2011}, to multi-partite entanglement \cite{SorensenPRL2001,KorbiczPRL2005,KorbiczPRA2006} and to 
other kinds of squeezing \cite{GuhnePR2009,BeduiniPRL2013}.  SSIs on symmetric states \cite{WangPRA2003} give the interesting result that spin squeezing implies entanglement of all pairs of atoms.  One set of SSIs, due to S\o{}rensen and M\o{}lmer \cite{SorensenPRL2001} and extended by Hyllus et al. \cite{HyllusPRA2012} quantifies the size of entangled groups based on the mean and variance of the collective spin.    These quantities are accessible in spin-squeezing experiments \cite{AppelPNAS2009,GrossN2010,RiedelN2010, LerouxPRL2010,ChenPRL2011,SewellPRL2012} and indeed are necessary to show metrologically-relevant spin squeezing \cite{WinelandPRA1992}.  The \SM~ SSI has been used to give indirect evidence for at least 80-partite entanglement  in a spin-squeezed Bose-Einstein condensate (BEC)  \cite{GrossN2010}.  This entanglement depth  far exceeds the current records of 8-partite entanglement in photons \cite{YaoNPhot2012} and 14-partite entanglement in ion chains \cite{MonzPRL2011}.  \btext{Recent work detecting multi-partite entanglement in non-polarized BECs \cite{LuckePRL2014} underscores the point that multi-partite entanglement is responsible for several kinds of non-classicality.}

%

}

{To date, no direct confirmation of these large entanglements, nor of any entanglement obtained by spin squeezing, has been demonstrated. Furthermore, various objections have been raised to the interpretation of squeezing in a BEC as highly-multipartite entanglement: It has been argued \cite{BenattiAoP2010} that the \SM~ results \cite{SorensenPRL2001}, which are based on statistics of distinguishable particles, are not applicable to experiments with indistinguishable particles, e.g. condensate atoms.  Prior to this, Hines et al. \cite{HinesPRA2003} argued ``the degree of entanglement between individual particles, unlike the entanglement between the modes, is not experimentally relevant so long as the particles remain in the condensed state.''  In contrast, Hyllus et al. \cite{HyllusPRL2010} describe an extension of the \SM~ results to indistinguishable particles, with the conclusion that BEC spin squeezing does imply multi-partite entanglement.   See also \cite{AmicoRMP2008} for a discussion in the context of quantum condensed matter.  A direct measurement of multi-particle states from a squeezed ensemble could help to settle these apparently conflicting claims.   

A direct test of the  squeezing-entanglement relation requires both the ability to generate squeezed states and to detect many individual particles in a state-selective way.  While atomic squeezing is available, the detection requirement is not presently met.  Detection of a large number of individual atoms in an optical lattice has been demonstrated \cite{NelsonNP2007,BakrN2009}, but at present this technique is not state-selective, except by the destructive method of removing atoms in all other states \cite{FukuharaNPhys2013}.  There is also a non-trivial challenge in combining squeezing and detection in a single atomic experiment.  

In this article we consider the prospects for observing highly multi-partite entanglement due to squeezing in a photonic system.  This is experimentally appealing, because both optical squeezing and state-selective multi-photon detection exist as mature technologies.   Combinations of squeezing and single-photon detection, e.g. \cite{WengerPRL2004,Neergaard-NielsenPRL2006} and multi-particle state-selective detection, e.g. \cite{KrischekNPhot2010,YaoNPhot2012} have been demonstrated in several labs.  Spin-squeezing theory for photonic systems has been less studied than for atoms \cite{BeduiniPRL2013}.  One major difference relative to atoms is the absence of superselection rules forbidding superpositions of different numbers of photons {\cite{DowlingPRA2006b}}.  Indeed, these superpositions are the hallmark of optical coherence \cite{GlauberPR1963,MolmerPRA1997,BartlettIJQI2006}.  We show how Glauber photodetection theory \cite{GlauberPR1963} can be used to describe the observable states in this scenario, and find that optical coherence provides an efficient route to highly multi-partite entanglement.  

We describe a simple scenario that produces polarization-squeezed light \cite{ChirikinQE1993,KlyshkoJETP1997,KorolkovaPRA2002,SchnabelPRA2003} by combining weakly squeezed vacuum with a coherent state \cite{PredojevicPRA2008}, and analyze various measures of multi-partite entanglement on the observable density matrix (ODM) from this state.   Extending the \SM~ + Hyllus et al. theory to large photon number, we find that the entanglement depth approaches a constant fraction of the total number of photons in the macroscopic limit.  We numerically evaluate witnesses for genuine multipartite entanglement (GME), as in \cite{JungnitschPRL2011,NovoPRA2013}, and find that all squeezed states are GME, a conclusion that is limited to $N=\NGME$ by computational power.  We evaluate $N$-partite entanglement by computing the reduced $2$-body density matrix, and find (again numerically) that states up to at least $N=100$ are $N$-partite entangled.  In all cases, trends suggest entanglement persisting also to higher numbers.  Concerning the feasibility, we show that entanglement can be observed in situations of significant loss and detector inefficiency, and without post-selection.  We find that the resource cost of both producing the squeezed state and detecting the entanglement therein scales as $N^{0}$, in stark contrast to most entanglement generation and detection.  Finally, we estimate that modern photon-counting imaging sensors could detect entanglement up to $N \approx 1000$.

}

\newcommand{\bp}{{\bf p}}

\newcommand{\D}{\hat{D}}
\newcommand{\Sq}{\hat{S}}
\newcommand{\detd}{d}
\newcommand{\detset}{D}

\newcommand{\subth}{_{\rm th}}

\newcommand{\nth}{n\subth}
\newcommand{\power}{q}
\newcommand{\no}[1]{\,:\!\!#1\!\!:\,\,}
\renewcommand{\choose}[2]{{ \left(\begin{array}{c} #1 \\ #2 \end{array} \right)}}

\begin{figure}[t]
\centering
\includegraphics[width=0.8 \textwidth]{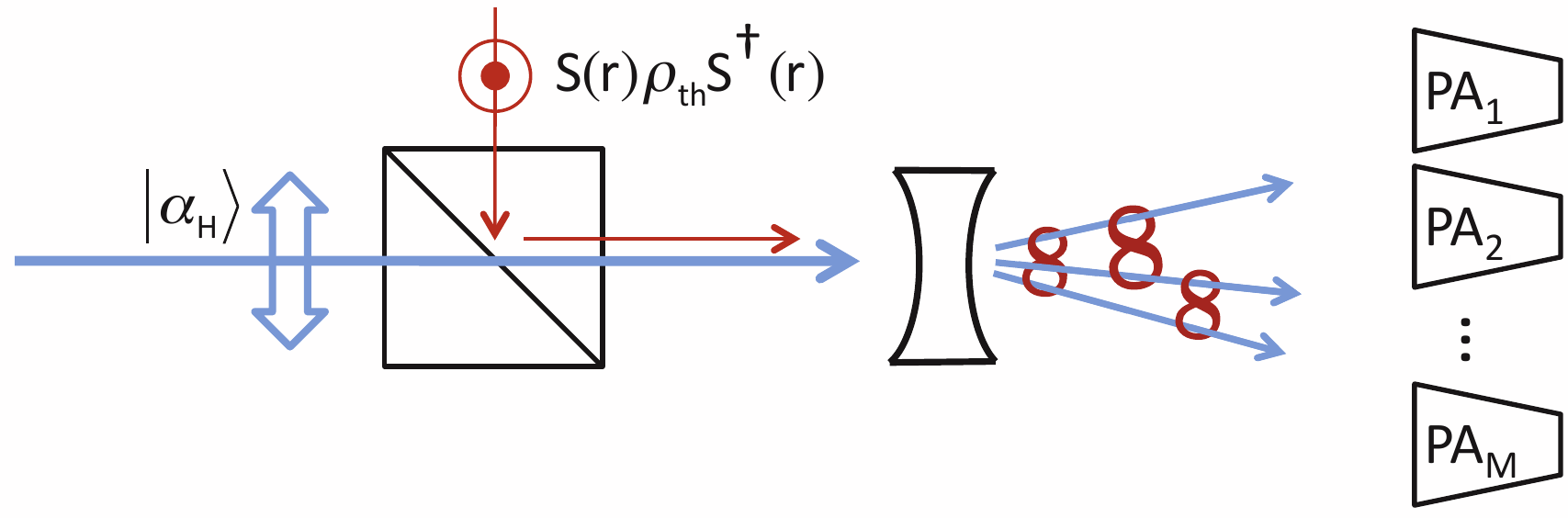}
\caption{Scenario for entanglement generation:  a horizontally-polarized coherent state $\ket{\alpha_H}$ is combined with a vertically-polarized squeezed state $S(r)\rho_{\rm th}S^\dagger(r)$.  The resulting polarization-squeezed state is symmetrically dispersed by a lens to an array of polarization analyzers (PA), each consisting of wave-plates, a polarizing beamsplitter, and single-photon detectors.  Under conditions typical of squeezing experiments, nearly all of the photons originate in the coherent beam and become polarization entangled via coherence with the much weaker squeezed beam. }
\label{fig:scenario}
\end{figure}

\newcommand{\asp}{A}
\PRLsection{Scenario}
\mtext{
We consider a single-mode pulse of polarization-squeezed light \cite{KorolkovaPRA2002,BowenPRL2002} described by the state
\be
\label{eq:RhoDef}
 \rho = \rho_H \otimes \rho_V
 \ee
consisting of a coherent state $\rho_H = \ket{\alpha_H}\bra{\alpha_H}$ in the horizontal ($H$) polarization and a squeezed thermal state $\rho_V = \Sq(r) \rho\subth \Sq^\dagger(r)$ in the vertical (V).  Here $\Sq^\dagger(r) a_V \Sq(r) = a_V \cosh r - a_V^\dagger \sinh r$ defines the squeeze operator $\Sq$ and $\rho\subth \equiv \sum_n {\nth^n}/{(1+\nth)^{n+1}} \ket{n}\bra{n}$ is the thermal state.  It is convenient to parametrize this state by $\nc \equiv |\alpha|^2$, $\ns \equiv \sinh^2 |r|$ and $\nth$.    States with this description can be produced by combining squeezed and coherent pulses on a polarizing beam-splitter \cite{PredojevicPRA2008} or by polarization self-rotation \cite{RiesPRA2003} of a coherent pulse.  {We assume this state can be prepared in a single spatial mode, an idealization which can be very well approximated by combining the two polarization modes in a single fiber. } Regarding $\rho_V$,  {for simplicity we take $r$ real and negative} and define $T^2 \equiv 1+2 \nth$, $\asp \equiv \sqrt{\ns} + \sqrt{\ns+1}$, so that the quadratures $x  \equiv (a_V+a_V^\dagger)/2$, $p \equiv i(a_V^\dagger - a_V)/2$ have variances $\var(x) = T^2 \asp/4$, $\var(p) = T^2/(4\asp^2)$.  Note that $p$ is  squeezed if $\ns >  \nth^2/(1+2\nth)$. 
}

 

\PRLsection{Macroscopic quantum features}
The quantum description of the macroscopic polarization is given by the Stokes operators \cite{SchnabelPRA2003}
\begin{eqnarray}
 S_0 &=& \frac{1}{2}(a^\dagger_H a_H + a^\dagger_V a_V) \\
 S_x &=& \frac{1}{2}(a^\dagger_H a_H - a^\dagger_V a_V) \\ 
 S_y &=& \frac{1}{2}(a^\dagger_H a_V + a^\dagger_V a_H) \\ 
  S_z &=& -\frac{i}{2}(a^\dagger_H a_V - a^\dagger_V a_H),
  \end{eqnarray}
from which
\begin{eqnarray}
2\expect{S_0} &=& \nc + \ns + \nth + 2 \ns \nth \\ 
2\expect{S_x} &=& \nc - \ns - \nth - 2 \ns \nth.
\end{eqnarray}
  If $\alpha=|\alpha|$ so that $S_z$ 
  is the lower-noise Stokes component, we find
   \begin{equation}
   4\var({S_z)}= \nc (1 + 2 \ns - 2 \sqrt{\ns(\ns+1)})(1 + 2 \nth).
   \end{equation} The state is squeezed 
 by the Wineland criterion \cite{WinelandPRA1994}, $\var(S_z) < \frac{1}{2}|\left<S_x\right>|$,  if { $\ns > \nth^2/(2\nth+1)$}, the same condition as for $\rho_V$ quadrature squeezing.  \btext{This form of the Wineland condition,  written for our situation in which the average polarization is along $S_x$, is a special case of a more general, SU(2)-invariant non-classicality condition relating the polarization components along the average and orthogonal directions \cite{LuisPRA2006}.  The entanglement results that follow are similarly invariant under global SU(2) rotations.  }

\begin{figure}[t]
\centering
\includegraphics[width=0.6 \textwidth]{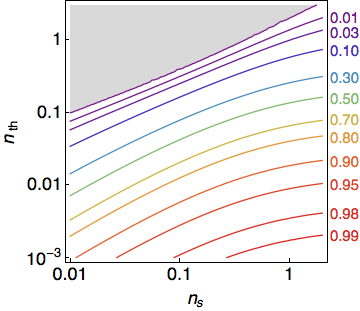}
\caption{{(color online)} {Highly-multipartite entanglement in weakly-squeezed macroscopic states.  Graph shows entanglement depth $k$ as calculated by the \SM~ + Hyllus theory, i.e. the lower bound to the size of entangled groups implied by the degree of polarization squeezing.   Contours show $ \lim_{S_0\rightarrow \infty} k/ (2\expect{S_0})$, i.e. entanglement depth as a fraction of total photon number in the macroscopic limit.  Axes show mean number of photons in the squeezed beam $\ns$ and mean number of thermal noise photons $\nth$, which parametrize the squeezing and mixedness of the state, respectively.  Contour values shown on right.   Grey region indicates non-entangled regime $k=1$, which coincides with non-squeezed states.  Results show that even for modest squeezing, entanglement depth is comparable to the number of particles.} }
\label{fig:LargeMS}
\end{figure}

\newcommand{\PNoise}{{\cal V}}
\newcommand{\PMag}{{\cal P}}
\newcommand{\delZ}{{\delta_Z}}
\newcommand{\JMS}{J_{\rm MS}}
\renewcommand{\JMS}{k}
\newcommand{\cV}{{\cal V}}
\newcommand{\cK}{{\cal K}}
\newcommand{\cZ}{{\cal Z}}
\newcommand{\cv}{{\cal v}}
\newcommand{\ck}{{\cal k}}
\newcommand{\cz}{{\cal z}}
\newcommand{\DMax}{d}

\PRLsection{Microscopic quantum features}
\subsection{Depth of entanglement}
To find the depth of entanglement $k$, i.e., the minimum size of entangled clusters in the state, we use the \SM~  theory 
which was shown by Hyllus {\em et al.} \cite{HyllusPRA2012} to apply also to indefinite numbers of \btext{degenerate} particles.  \btext{We apply the Hyllus et al. version of the \SM~theory to the polarization-squeezed beam before dispersion to the detector array.  Being a degenerate ensemble of two-state particles sharing a single external mode, this is precisely the situation discussed by Hyllus et al.}
 {In the Appendix, we describe the macroscopic limit of this theory and apply it to the polarization-squeezed state described above.  We find that 
 $k$ is proportional to the total number of photons $2\expect{S_0}$, implying a macroscopic entanglement depth.  The entanglement depth fraction $k/(2\expect{S_0})$ can approach unity for small $\nth$.  See Figure \ref{fig:LargeMS}.}

\newcommand{\RNorm}{{\cal R}}
\newcommand{\ntot}{{n_{\rm tot}}}
\renewcommand{\ntot}{{n_{T}}}
\newcommand{\supN}{^{(N)}}
\newcommand{\supntot}{^{(\ntot)}}
\newcommand{\subntot}{_{\ntot}}

\subsection{State of detected $N$-tuples}

{We consider now the polarization state of $N$ photons detected from the state $\rho$.  Concretely, we imagine passively splitting the beam to a large number $M \gg N$ of analyzers, as illustrated in Figure \ref {fig:scenario}, and using the observed detections to infer the locations of the photons, as well as to measure their polarization.   By permutational invariance and symmetry of the splitting process, the detected state depends only on $N$ and $\rho$.  The same ``observable density matrix'' (ODM) $\RNorm^{(N)}$ describes the photons independently of which detectors they arrive to.  \qtext{A possible detection system would use liquid-crystal spatial light modulators (SLMs) to affect independent polarization rotations \cite{EriksenPC2001,MorenoOE2012} and thus specify $M$ independent polarization bases, a large polarizing beamsplitter (PBS) to simultaneously split all the $M$ channels, and one electron-multiplied charge-coupled device (EMCCD) imaging sensor at each PBS output, to serve as $M$ independent photon counters.}



\btext{
To compute $\RNorm^{(N)}$, we follow standard methods \cite{JamesPRA2001} based on the Glauber theory of photodetection \cite{GlauberPR1963}.  
}
For each analyzer we define a detection mode,  described by the annihilation operator $a_{d,p} =  \sqrt{M^{-1}} a_{p}  + \sqrt{1-M^{-1}}  a^{({\rm aux})}_{d,p} $ where $d$ is the analyzer index, $p$ is the polarization and $a^{({\rm aux})}_{d,p}$ describes an auxiliary mode, required for unitarity of the splitting process and assumed to be in a vacuum state.  
 
\newcommand{\bpi}{{{\bm \pi}}}
\newcommand{\kpis}[1]{\ket{{\bm \pi}_{#1}}}
\newcommand{\bpis}[1]{\bra{{\bm \pi}_{#1}}}

For a single pulse, the {average number of coincidence detections} with polarizations $\bp^{(1)},\ldots,\bp^{(N)}$ at detectors $1\ldots N$ is given by 
\bea
\label{eq:GlauberPDef}
P^{(N)}_{\{\bp\} } &\propto& \expect{a^\dagger_{1,\bp^{(1)}} 
\ldots a^\dagger_{N,\bp^{(N)}}  a_{N,\bp^{(N)}} \ldots 
a_{1,\bp^{(1)}}} \nnpropto
\expect{a^\dagger_{\bp^{(1)}} 
\ldots a^\dagger_{\bp^{(N)}}  a_{\bp^{(N)}} \ldots 
a_{\bp^{(1)}}},
\eea 
where $\expect{\cdot}$ indicates expectation with respect to $\rho$.  {In our regime, with $M$ much larger than the average number of photons in a pulse, the probability of an $N$-fold detection at any given group of $N$ detectors is $\ll 1$, and $P^{(N)}_{\{\bp\} }$ is, for all practical purposes, the probability of this $N$-fold coincidence.} The second line follows from the definition of $a_{d,p}$ and the fact that the vacuum contributes nothing to the normally-ordered expectation.  
We employ the computational basis $\kpis{0}\equiv \ket{H,\ldots,H,H},  \kpis{1}\equiv \ket{H,\ldots,H,V}, \ldots , \kpis{2^N-1}\equiv \ket{V,\ldots,V,V}$ to describe the (unnormalized) ODM
\bea
\label{eq:GlauberRDef}
R^{(N)}_{\kpis{i} ,\bpis{j}} &\propto& \expect{a^\dagger_{{\bm \pi}_{i}^{(1)} }
\ldots a^\dagger_{{\bm \pi}_{i}^{(N)}}  a_{{\bm \pi}_{j}^{(N)}} \ldots 
a_{{\bm \pi}_{j}^{(1)} } },
\eea
where ${\bm \pi}_{i}^{(l)}$ indicates the polarization of the $l$'th photon in the basis element $\kpis{i}$. {The elements of $R^{(N)}$ can be directly calculated using phase-space techniques (see Appendix).  Most elements of $R^{(N)}$ are not directly measurable, but can tomographically reconstructed from $P^{(N)}_{\{\bp\} }$. The detection probabilities are given by the Born-rule-like relation
\be
P^{(N)}_{\{\bp\} } \propto {\rm Tr}[ \Pi_{\vec{p}} R^{(N)} ],
\ee
where $\vec{p}$ is defined to give  $\bp^{(i)} =  \sum_{j} p_j \bpi_{j}^{(i)}$, so that   
 $\Pi_{\vec{p}}  \equiv  \vec{p} \wedge \vec{p}$ is a projector onto $\{\bp\}$.   For example, for $N=2$, the computational basis is
 $\kpis{0}= \ket{H,H},  \kpis{1}= \ket{H,V},  \kpis{2}= \ket{V,H}$ and $\kpis{3}= \ket{V,V}$.  Detection of polarizations $H,V$ at detectors $1,2$, respectively, is described by  $\{ {\bf p} \} = ({H},{V})$ and thus $\vec{p} = (0,1,0,0)$.  ${\rm Tr}[ \Pi_{\vec{p}} R^{(N)} ]$ is then the $2,2$ element of $R^{(N)}$.   
 Finally, we normalize $\RNorm^{(N)} \equiv R^{(N)}/{\rm Tr}[R^{(N)}]$. 
}

\btext{
We note aspects of the detection process that can reduce the coherence of $\RNorm^{(N)}$ relative to $\rho$.  First, $\rho$ describes a superposition of different total photon number, and the coherence between different numbers is lost in photon counting.   Second, $\RNorm^{(N)}$ is a reduced density matrix, obtained by tracing over the undetected photons.  Third,  parts of $\rho$ with different photon number can contribute to $\RNorm^{(N)}$, making $\RNorm^{(N)}$ a mixture of these different contributions.  Photon entanglement in $\RNorm^{(N)}$  implies photon entanglement in $\rho$, because these transformations by which $\rho$ gives $\RNorm^{(N)}$, i.e. decoherence, reduction, and mixing, do not produce entanglement.  
}

\newcommand{\mNFactor}[2]{{F_{#1,#2}}}
\newcommand{\mNFactorG}[2]{{G_{#1,#2}}}
\newcommand{\post}{{\rm PS}}
\newcommand{\NPost}{{N_{\rm P}}}
\newcommand{\dets}{{\{k\}}}
\newcommand{\equalsets}{=}

\newcommand{\diffnoise}{\xi}
\newcommand{\Nx}{\diffnoise_x}
\newcommand{\Np}{\diffnoise_p}
\newcommand{\zp}{\zeta_p}
\newcommand{\zx}{\zeta_x}
\newcommand{\zpm}{\zeta_\pm}
\newcommand{\loss}{{\rm loss}}
\newcommand{\byloss}{\stackrel{\loss}{\rightarrow}}

\newcommand{\hs}{\hermsym}

\subsection{Illustration: ODM for $N=2,3$}
$\RNorm^{(N)}$ is in general a mixture of symmetric states, e.g. $\RNorm^{(2)}$ is a mixture of $\ket{H,H} \pm \ket{V,V}$,  $ \ket{H,V} + \ket{V,H}$, and either $\ket{H,H}$ or $\ket{V,V}$.  We show $\RNorm^{(2)}$ and  $\RNorm^{(3)}$ as examples. 
\be
\label{eq:rhotwo}
\RNorm^{(2)} \propto
\left(
\begin{array}{cccc}
A_2 & \hermsym & \hermsym & D_2 \\
  & B_2 & B_2 & \hermsym \\
  &    &  B_2 & \hermsym \\
  \makebox[0mm]{\vspace{-0mm}\hspace{+5mm}h.c.} &  &  & C_2 \end{array}
\right)
\ee
where the basis is $\ket{H,H},\ket{H,V},\ket{V,H},\ket{V,V}$ and $A_2 \equiv \nc^2$, $B_2 = \nc (\ns + \nth + 2 \ns \nth)$, $C_2 = 3 \ns^2 (1+ 2 \nth)^2  + \ns (1 + 8 \nth + 12 \nth^2)+  2 \nth^2$, and $D_2 =  \nc \sqrt{{\ns(\ns+1)}}(1+2 \nth)$.
The symbol ``$\hs$'' indicates a zero entry, while ``h.c.'' indicates the lower-triangular portion of the (hermitian) matrix.  In the limit of zero noise $\nth \rightarrow 0$, and small flux $\ns = {\nc^2} \rightarrow 0$, $\RNorm^{(2)}$ describes a Bell state $\propto \ket{HH}+\ket{VV}$.

Similarly,
\be
\label{eq:rhothree}
\RNorm^{(3)} \propto
\left(
\begin{array}{cccccccc}
A_3 &\hs &\hs & E_3 &\hs & E_3 & E_3 &\hs \\
   & B_3 & B_3 &\hs & B_3 &\hs &\hs & F_3 \\
   &  & B_3 &\hs & B_3 &\hs &\hs & F_3  \\
  &  &  & C_3&\hs & C_3 & C_3 &\hs \\
  & & &   & B_3 &\hs &\hs & F_3 \\
  &    & \makebox[0mm]{h.c.}   & &   & C_3 & C_3 &\hs \\
  &    &   & &   & & C_3 &\hs \\
   &  &  &   & &  &   & D_3
\end{array}
\right)
\ee
where 
\bea
A_3 &=&   \nc^3 \\  
B_3 &=&   \nc^2 (\ns + \nth + 2 \ns \nth) \\
 C_3 &=& \nc [3\ns^2(1+2\nth) + \ns(1+8\nth + 12 \nth^2)+2 \nth^2] \nn \\
  D_3 &=&  3 (\ns + \nth + 2 \ns \nth) [2 \nth^2 \nnp \ns (1 + 2 \nth) (3 + 10 \nth) + 
   5 (\ns + 2 \ns \nth)^2] \\
     {E_3} &=&    \nc^2 \sqrt{\ns(1 + \ns)} (1 + 2 \nth)  \\
      F_3 &=&  3 \nc \sqrt{\ns(1 + \ns)} (1 + 2 \nth) (\ns + \nth + 2 \ns \nth).
   \eea  In the same limit, $\RNorm^{(3)}$ describes a state $\propto \ket{HHH} + \ket{HVV} + \ket{VHV} + \ket{VVH}$ with maximal 3-tangle \cite{CoffmanPRA2000}.

\newcommand{\xmin}{x_{\rm min}}

\begin{figure}[t]
\centering
\includegraphics[width=0.5 \textwidth]{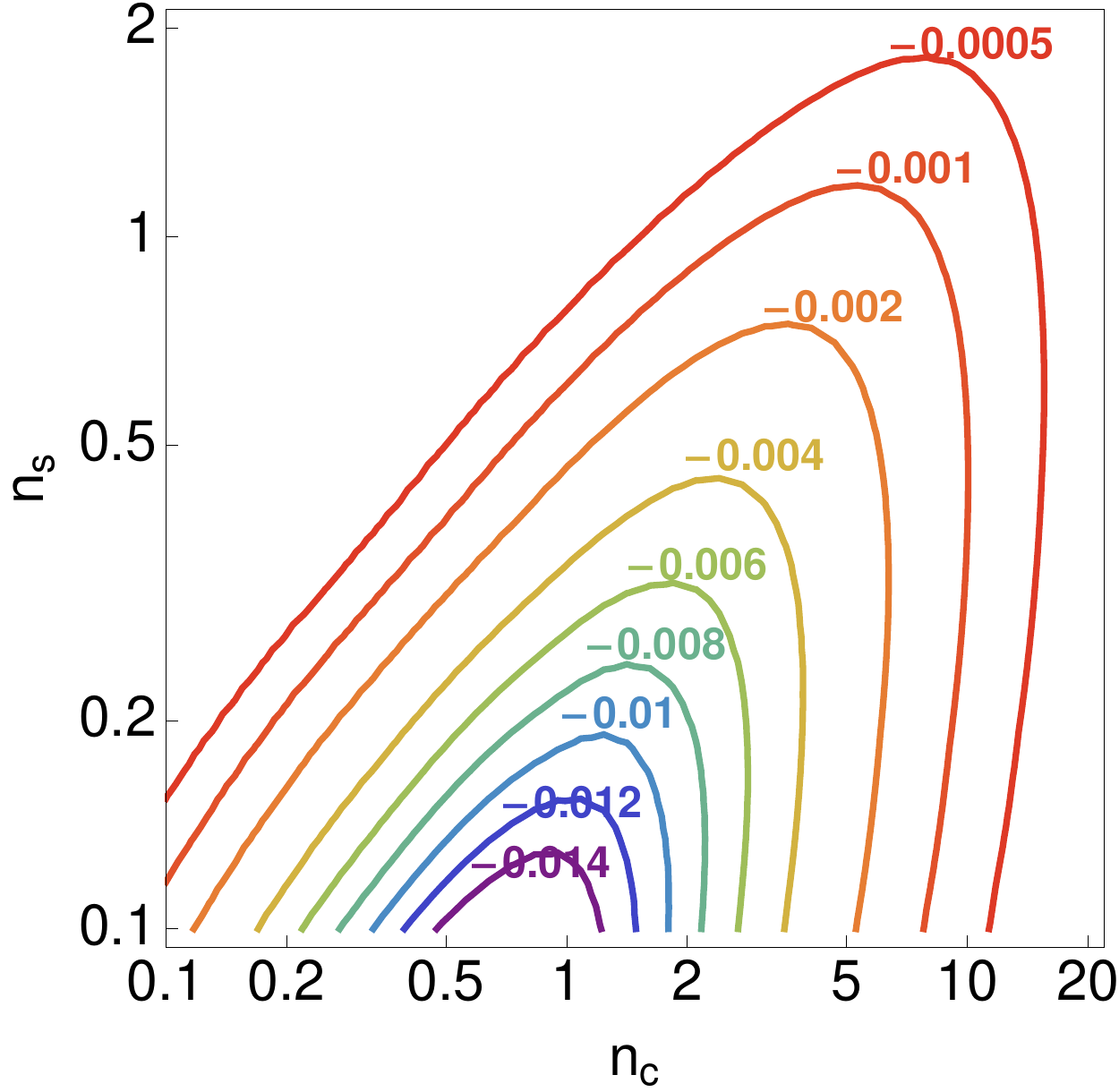} 
\includegraphics[width=0.5 \textwidth]{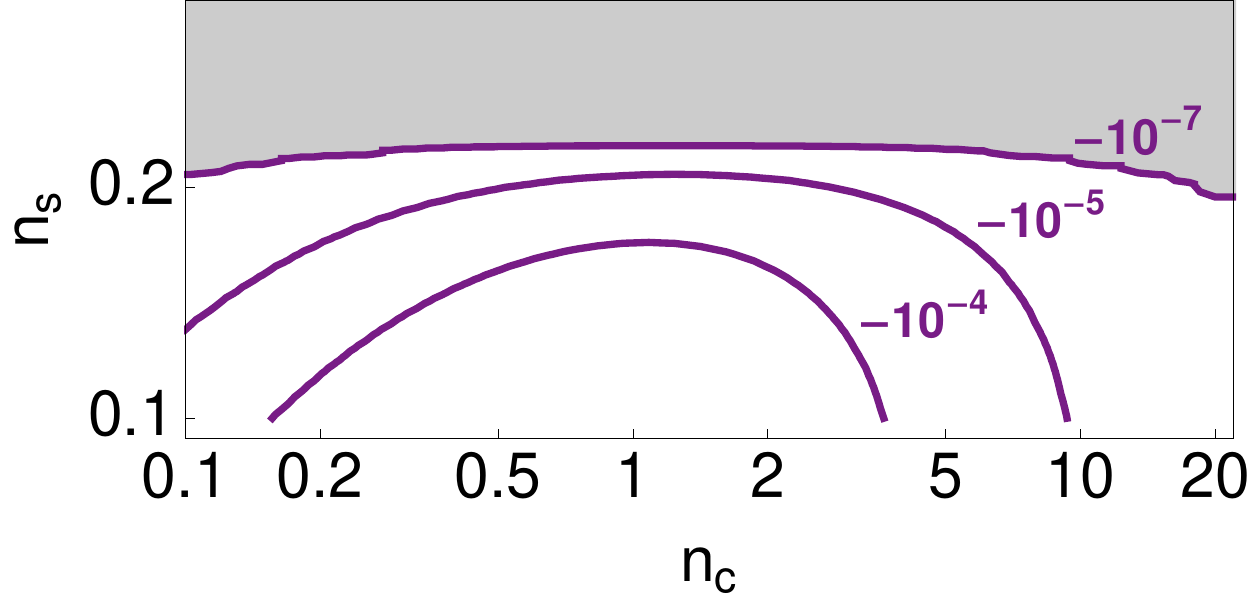} 
\caption{{(color online)} Genuine multipartite entanglement of the $N=4$ state.  Minimal expectation value $\xmin \equiv \min_W \expect{\RNorm^{(N)} W}$ with respect to all fully decomposable witnesses $W$, computed for $N=4$ states with $\nth = 0$, and varying $\nc,\ns$.   {Upper graph:} For all the $\nc, \ns$ tested, $\xmin < 0$, indicating genuine multipartite entanglement of these states \cite{NovoPRA2013}.     {Lower graph: } $\xmin$ computed on the $N=4$ state after averaging over relative optical phase between the $H$ and $V$ parts of the state.   This decoherence operation reduces but does not fully eliminate the GME. Grey region indicates {$\xmin < 10^{-7}$ (zero, to within computational uncertainties),} i.e. non-GME states. }
\label{fig:GME} 
\end{figure}

\begin{figure}[t]
\centering
\includegraphics[width=0.8 \textwidth]{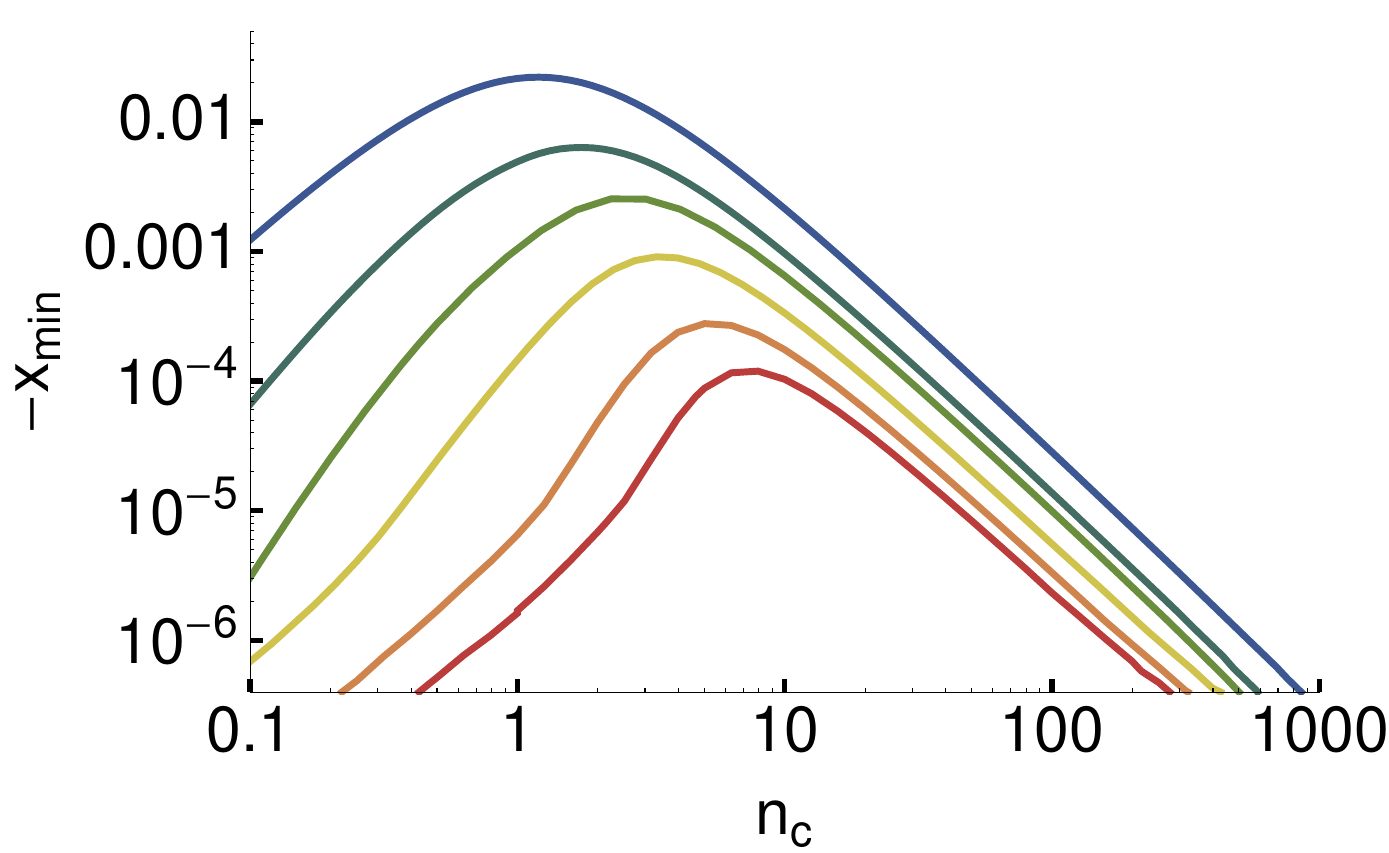}
\caption{{(color online)} Negativity -$\xmin$ as in Fig. \ref{fig:GME} versus $\nc$ with fixed $\ns=0.3$.  Different curves show, from top to bottom, $N=3,4,5,6,8,10$.  
Although memory constraints limited the calculation to $N \le \NGME$, the trend suggests that negativity  persists also to larger $N$.  }
\label{fig:GMETrend} 
\end{figure}

\subsection{Observable state in terms of pure states} 
\mtext{We can simplify calculation of  $\RNorm^{(N)}$ by considering the effect of polarization-independent losses.  If $a_{i,\pi} \byloss  a_{i,\pi} \sqrt{\eta} + a_{i,\pi}^{({\rm res})} \sqrt{1-\eta}$ where $\eta$ is the transmission and $a_{i,\pi}^{({\rm res})}$ describe reservoir modes, assumed to contain vacuum, then 
$R^{(N)}  \byloss \eta^N R^{(N)}$ and $\RNorm^{(N)}$ is unchanged.  {This is a well-known property of coincidence detection: global losses change only the rate of coincidences, not the post-selected quantum state. } At the same time, $\nc \byloss \eta \nc$, and $\var(q) \byloss \eta \var(q)  + (1-\eta) /4$, where $q \in \{x,p\}$, from which we find that any mixed squeezed state $\rho$, with $\nc$, $\ns$, $\nth$, can be produced from a  pure state $\rho'$ with $\nc' =  \nc /\eta$, $\ns' =(\ns + 2 \ns \nth+ \nth )/\eta$ and $\nth'= 0$ for  $\eta = (\ns + 2 \ns \nth - \nth^2)/(\ns + 2 \ns \nth + \nth)$.} 
{From here on, we consider only $\nth=0$ states, with the understanding that the $\nth \ne 0$ results can be obtained by the above transformation.  }

{
\subsection{Genuine multipartite entanglement }
Multipartite states which cannot be decomposed into mixtures of bi-separable states (under any partition) are called genuine multipartite entangled.  
Using code by  J\"{u}ngnitsch {\it et al.} \cite{JungnitschPRL2011}, we find that $\RNorm\supN$ is genuine $N$-partite entangled, within our numerical ability to test.  With $N$ up to 5, we use the approach of positive partial transpose mixtures, and compute the minimal expectation value $\xmin \equiv \min_W \expect{\RNorm^{(N)} W }$ where the minimization is  with respect to all fully decomposable witnesses $W$ (function {\tt fdecwit}).  If $\xmin$ is negative, $\RNorm^{(N)}$ is genuine multipartite entangled \cite{NovoPRA2013}.  As illustrated in Fig. \ref{fig:GME}, $\RNorm^{(N)}$, with $N=3,4,5$, was tested over a broad range of $\nc,\ns$, and always found to be GME.   In Fig. \ref{fig:GMETrend} we show $\xmin$ computed with fixed $\ns = 0.3$ and $\nc$ up to 1000.   The trends in $\xmin$ clearly suggest that GME should persist for large $N$, but memory limitations prevented testing this suggestion for $N > 5$. 
}

\subsection{Multipartite entanglement}

Simple $N$-partite entanglement is easier to show than GME, and here we are able to show $N$-partite entanglement up to $N=100$, again numerically.  Here it suffices to show that any pair of photons is entangled.
We study the reduced two-photon density matrix, obtained by tracing over  any $N-2$ photons 
\be
\RNorm^{(2,N)}  \equiv {\rm Tr}_{N-2} [\RNorm^{(N)}].
\ee
%
By permutation symmetry, an entangled $\RNorm^{(2,N)}$ implies that no partition of $\RNorm^{(N)}$ is separable.  As shown in the Appendix, $\RNorm^{(2,N)}$ can be computed in $O(N^3)$ time using phase-space methods.  

\newcommand{\Cmono}{{\cal C}_{\rm max}}
\begin{figure}[t]
\centering
\includegraphics[width=0.8 \textwidth]{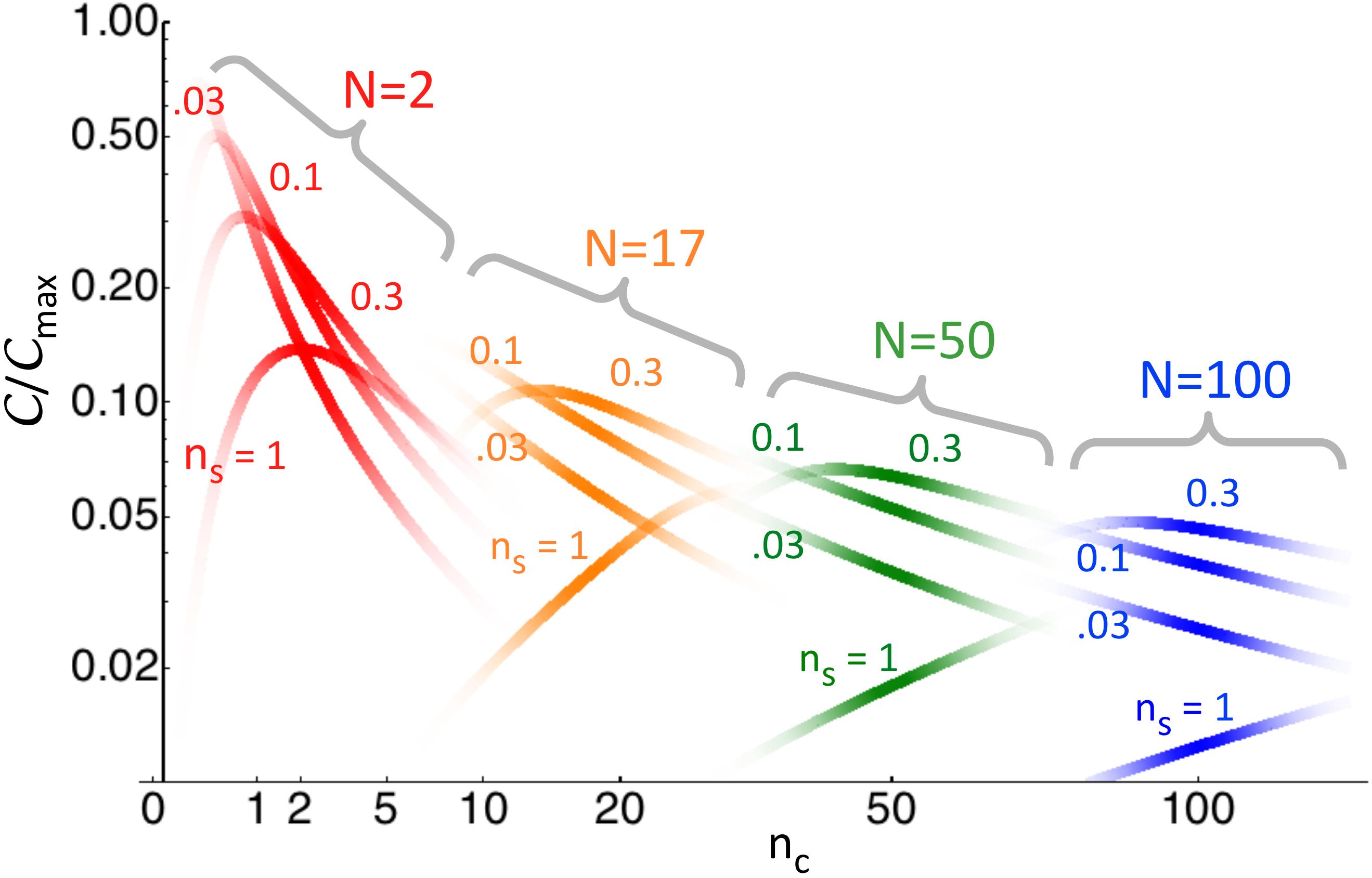}
\caption{{(color online)} Scaled concurrence ${\cal C}/\Cmono$ as a function of  $N$, $\nc$, and $\ns$ for pure states, i.e., $\nth = 0$.  Here $\Cmono = 1/\sqrt{N-1}$ is the largest concurrence allowed by entanglement monogamy \cite{WoottersPRL1998}, shown for $N=2$ (red),  $N=17$ (orange),  $N=50$ (green) and $N=100$ (blue).  Labels indicate $\ns$ values, and line strength indicates the likelihood of observing $N$ photons from the coherent state with $\nc$ average photons, calculated as $P_N(\nc)/P_N(N)$, where $P_N(\lambda)$ is the Poisson distribution.}
\label{fig:ConcurrenceForDifferentN} 
\end{figure}

As with GME, we find that, to within computation limits, any squeezed state $\RNorm^{(N)}$ is $N$-partite entangled.  
For pure states, i.e., $\nth=0$, we compute the observable concurrence ${\cal C}$, i.e., the concurrence computed on $\RNorm^{(2,N)}$, shown in Figure \ref{fig:ConcurrenceForDifferentN}.  In all cases, Wineland-criterion squeezing $\ns>0$ gives nonzero concurrence. For example, with $\nc=100,$ $\ns=0.3$, and $N=100$,
\be
\label{eq:rhotwoN}
\RNorm^{(2,100)} \approx
\left(
\begin{array}{cccc}
0.9440 & \cdot & \cdot & 0.0293 \\
\cdot  & 0.0270 & 0.0270 & \cdot \\
\cdot  &   0.0270 &  0.0270 & \cdot \\
0.0293 & \cdot & \cdot  & 0.0021 \end{array}
\right). 
\ee
The concurrence is ${\cal C} \approx 0.00468$, which is 4.7\% of that allowed by monogamy \cite{WoottersPRL1998}, i.e., $\Cmono = 1/\sqrt{N-1} \approx 0.10$.

{
\subsection{Role of optical coherence}
We note that the entanglement properties are largely due to optical coherence between the $H$- and $V$-polarized parts of the state in Eq. (\ref{eq:RhoDef}).  By optical coherence, we understand the non-vanishing of phase-dependent correlation functions such as $\expect{a_H}, \expect{a_V}$ and $\expect{a^\dagger_H a_V}$. These are not invariant under the phase rotations $U_H(\phi_H) \equiv \exp[i \phi_H a_H]$ and $U_V(\phi_V) \equiv \exp[i \phi_V a_V]$.  We can see the role of optical coherence by analyzing the decohered state $\expect{ U_H(\phi_H) \rho U_H^\dagger(\phi_H)}_\phi$, where $\expect{\cdot}_\phi$ indicates a statistical average over $\phi$.  This decoherence removes most, but not all of the non-classical features described above:  it mixes the quadratures $x$ and $p$, and thus averages their variances, leaving no squeezing.  Without squeezing, the \SM~ + Hyllus theory is mute: no entanglement is implied.  Decoherence sets the off-diagonal element $\RNorm^{(2,N)}_{1,4}$ to zero, because this arises from states differing by two in number of $H$ photons.  This decohered $\RNorm^{(2,N)}$ was tested for multi-partite entanglement, using the same ranges as in Fig. \ref{fig:ConcurrenceForDifferentN}, and none was found. In contrast, testing the decohered $\RNorm^{(N)}$ over the same range as in Fig. \ref{fig:GME}, the decohered state is in fact GME for $\ns < 0.25$, although at a lower level than for the original state.  For example, with $\nc = 1, \ns = 0.1$, the optimal witness finds $\xmin = -1.5 \times 10^{-3}$, versus $-1.5 \times 10^{-2}$ for the non-decohered state.   See inset of Fig. \ref{fig:GME}.  This entanglement evidently arises from the phase-independent off-diagonal terms.}

\subsection{deterministic entanglement}
$\RNorm^{(2,N)}$ does not depend strongly on $N$, and the likely outcomes
$ \nc - \sqrt{\nc}\le N\le \nc + \sqrt{\nc}$ have similar ${\cal C}/ {\cal C}_{\rm max}$. 
{For illustration, we compute the reduced ODM averaged over the possible outcomes $N$ as
 $\RNorm^{(2,\expect{N})} \equiv \sum_N P_N \RNorm^{(2,N)}$, where $P_N$ is the probability of observing $N$ photons.  If $\RNorm^{(2,N)}$ varies strongly with $N$, this mixing of different contributions will reduce coherence. 
 In the simplest case, i.e. $\nth = 0$ and $\eta = 1$, $P_N$ is given by the convolution of 
the coherent state number distribution $P_{N_c} = \exp[-\nc] \nc^{N_c}/{N_c}!$ and the squeezed vacuum number number distribution $P_{N_s} = \ns^{N_s} (2 N_s)! / [4^{N_s} (1+\ns)^{-N_s-1/2}(N_s!)^2]$.  
The resulting state, as above with $\nc = 100$ and $\ns = 0.3$, is
\be
\label{eq:rhotwoNbar}
\RNorm^{(2,\expect{N})} \approx
\left(
\begin{array}{cccc}
0.9336 & \cdot & \cdot & 0.0333 \\
\cdot  & 0.0310 & 0.0310 & \cdot \\
\cdot  &   0.0310 &  0.0310 & \cdot \\
0.0333 & \cdot & \cdot  & 0.0030 \end{array}
\right)
\ee
with concurrence ${\cal C} \approx 0.00464$.  This only slightly less than the most probable outcome $N=100$ shown in Eq. (\ref{eq:rhotwoN}).  
We conclude that post-selection of a particular $N$ is not necessary to observe the two-body correlations that imply entanglement.  
}

For these likely states, ${\cal C}$ peaks for $\ns \approx 0.3$, or $\approx 4.5$ dB of squeezing. 
{We conjecture that this is a manifestation of entanglement monogamy, in that a small amount of squeezing produces mostly \btext{concurrence in the reduced two-body density matrix, a measure of} bipartite entanglement, and that larger squeezing produces {increasing \btext{measures} of higher-partite entanglement,} \btext{for example more 3-tangle in the reduced three-body density matrix}.  By entanglement monogamy, these higher entanglements come at the expense of bipartite entanglement.  It would be interesting  to compute the 3-tangle on these states to test this conjecture \cite{CaoPRA2010, MoroderPersComm}.}

\PRLsection{Prospects for observation}
Due to its structure, $\RNorm^{(2,N)}$ is negative under partial transpose if $\Delta \equiv |\RNorm_{1,4}^{(2,N)}| - \RNorm_{2,3}^{(2,N)} > 0$, e.g. if $0.0293 > 0.0270$ in $\RNorm^{(2,100)}$ above.  For  $\nc=N$ in the range $2$ to $100$ and $\ns$ chosen to maximize ${\cal C}$, we find $\Delta = 0.23/N$ to within 10\%.  \mtext{Although reconstruction of general $N$-particle states requires  resources exponential in $N$, here the method of preparation ensures permutation invariance (PI), allowing use of PI reconstruction methods to avoid exponential costs \cite{TothPRL2010} .  In particular, we can find $R^{(2,N)}$ by averaging all pairs of detected photons, i.e., $N(N-1)$ pairs {per shot}, giving single-shot statistical errors of $\delta \Delta \sim 1/N$.   We then need $O(N^0)$ shots to establish $\Delta > 0$ with unit signal-to-noise.  }

\mtext{

In light of the $N^0$ scaling of both production and characterization of entanglement,  $N$ is only limited by the number of available detectors.  {In this regard, EMCCD sensors are promising:} They can have $M = 2^{20} > 10^6$ pixels, 90\% quantum efficiency, negligible dark-count rates, single-photon sensitivity, and some degree of photon-number discrimination for $<5$ photons at a given pixel \cite{MackaySPIE2010}.  If we limit ourselves to $N^2 < {M}$, fewer than one pixel on average receives multiple photons, so that photon-number resolution is not required \btext{(a simple calculation of ${\rm Tr}[ R^{(2)}]$, i.e., of the un-normalized ODM, shows that bunching is not significant)}.  In this scenario, and with $M=2^{20}$,  {entangled states with $N\sim 1000$} are detectable.  {We leave for future work the question of whether photon-number resolution could allow these sensors to measure entanglement in states with $N > M^{1/2}$.}
%
}

%


\PRLsection{Conclusion}
{ In analogy with ``extreme spin squeezing,'' in which squeezing of collective spin variables implies highly multi-partite entanglement of the constituent particles, we have quantified the photonic entanglement of a practical polarization-squeezed state.  We first adapt the S\o{}rensen-M\o{}lmer-Hyllus theory to large particle number, and find that Wineland-criterion squeezing implies entanglement of a macroscopic fraction of the total photons present.  We describe the observable $N$-photon density matrix in terms of Glauber photodetection theory, and find that the Wineland criterion similarly implies genuine $N$-partite entanglement, for $N$ up to at least 10 (limited by computational power), and by a less stringent criterion $N$-partite entanglement up to at least $N=100$.  This entanglement is considerable even for modest squeezing and highly robust against losses.  The resource cost of producing and detecting these states scales as $N^0$, leaving the number of detectors as the limiting factor for size of the observable entangled states.  We estimate that with available detectors $\sim$1000-partite entanglement should be observable. 
} 

{These results provide the first practical route to a test of ``extreme spin squeezing,'' \cite{SorensenPRL2001} in which squeezing of macroscopic quantum variables produces highly multi-partite entanglement.  The experimental study of this subject may answer open fundamental questions about the nature of entanglement in systems of indistinguishable particles \cite{HinesPRA2003, WisemanPRL2003, GrossN2010, HyllusPRL2010}, the role of entanglement in condensed matter systems \cite{DowlingPRA2006, AmicoRMP2008} and quantum simulation of exotic phases of many-body systems \cite{TagliacozzoNC2013,HauckePRA2013}.  The highly-entangled states described here may also have application in quantum networking: atom-resonant polarization-squeezing \cite{PredojevicPRA2008} combined with heralded absorption of photons \cite{PiroNP2011,SpechtN2011} could be used to transfer the photonic entanglement onto atomic registers.}



~ 

\ack
\btext{The question of whether polarization-squeezed light might consist of entangled photons was provoked by Ref. \cite{MolmerPRA1997} (see also \cite{BartlettIJQI2006}).}  We thank P. Hyllus, J. K. Korbicz, G. Messin, G. Toth, R. Sewell, L. Tarruell and A. S{\o}rensen for helpful discussions.  Zhen Wang and Lixin He graciously contributed the 3-tangle of $\RNorm^{(3)}$ at an early stage.  Otfried G\"{u}hne and L. Novo graciously shared with us their PI GME code from \cite{NovoPRA2013}.  Work was supported by the Spanish MINECO  project MAGO (Ref. FIS2011-23520) and by the European Research Council project AQUMET.

\PRLsection{additional materials}
{The calculations of this manuscript are available as a Mathematica notebook, in an ancillary file at \href{http://arxiv.org/abs/1304.2527}{http://arxiv.org/abs/1304.2527}}
~ 

\appendix
\PRLsection{Large-J S\o{}rensen-M\o{}lmer-Hyllus theory} 
{
\SM~+Hyllus theory \cite{SorensenPRL2001,HyllusPRL2010} starts from the observation that the possible states of a spin-$J$ system with spin operator ${\bf j}$ are bounded in the space of scaled spin polarization $\zeta \equiv \expect{j_z}/J$ and  scaled spin noise $\upsilon \equiv \expect{j_x^2}/J$.   The range of allowed $(\upsilon,\zeta)$ increases with $J$ so that any given $(\upsilon,\zeta)$ implies a minimum $J$.  By a convexity argument, an ensemble of $N_A$ non-entangled  spin-$J$ systems is  bounded by the same curve in the space $(\cZ \equiv \expect{J_z}/[N_A J], \cV \equiv \expect{J_x^2}/[N_A J] )$ where ${\bf J} \equiv \sum_i {\bf j}^{(i)}$ is the collective spin of the ensemble.  Consequently, a given $(\cV,\cZ)$ implies a minimum $J$ in the ensemble.  If this spin-$J$ system is composed of spin-1/2 particles, it implies the presence of a group of at least $2 J$ entangled particles.  This number $k \equiv 2J$ is known as the depth of entanglement.

The bound on $(\cV,\cZ)$ is found by minimizing $\expect{j_H} \equiv \expect{j_x^2 - \mu j_z}$ over possible spin-$J$ states to find a family of optimal states parametrized by $\mu$.  This was done for small $J$ by numerical matrix diagonalization in \cite{SorensenPRL2001}.  
For highly polarized large-$J$ states, we can approximate  the spin raising operator $j_+ \ket{J,m} = \sqrt{J(J+1)-m(m+1)} \ket{J,m+ 1} $, where $m$ is the magnetic quantum number,
  as $j_+ \approx b \sqrt{2 J +1}$ where $b$ is the annihilation operator for the defect $\delZ \equiv J-m$, i.e., if $\ket{\delZ} = \ket{J,J-\delZ}$, then $b \ket{\delZ} = \sqrt{\delZ} \ket{\delZ-1}$.  This geometrical change neglects terms of order 
 $\sqrt{\delZ/J}$
   \cite{ArecchiPRA1972}.

In terms of $j_+$ and $j_- \equiv j_+^\dagger$, we have 
$j_x^2 = (j_+ + j_-)^2/4 \approx (2J+1)(b+ b^\dagger)^2/4$, $j_z \approx J - b^\dagger b$, and thus
$j_H \approx (2J+1)(b+ b^\dagger)^2/4 + \mu b^\dagger b - \mu J$. 
%
\newcommand{\bdelZ}{\bar{\delta}_Z} 
This operator is quadratic in $b,b^\dagger$ and thus reducible to
$j_H = A c^\dagger c + B$ by a Bogoliubov transformation $b \rightarrow c \cosh r + c^\dagger \sinh r$.  $\expect{j_H}$ is thus minimized by a state obeying $c\ket{\phi}=0$, i.e., squeezed vacuum in the $b$ picture.  Squeezed vacuum shows $\expect{(b+b^\dagger)^2} = \exp[-2r]$ and $\bar{\delta}_Z \equiv \expect{\delZ}   = \expect{b^\dagger b} = \sinh^2 r$, from which $2 \expect{j_x^2}/J \approx 1+2 \bdelZ - 2 \sqrt{\bdelZ(1+\bdelZ)}$, after dropping a term of order $1/J$  \footnote{ A numerical calculation, using $j_H$ truncated to the subspace $J-\DMax \le m \le J$ and increasing $\DMax$ until convergence, agrees with this approximation with an error in $\expect{j_x^2}$ of at most $5\sqrt{\bdelZ}/(2J)$  for $0 \le \bdelZ \le 2$, corresponding to up to 10 dB of squeezing. }.  In terms of $(\upsilon, \zeta)$, we have $2\upsilon = 1+2 J (1-\zeta) - 2 J \sqrt{(1-\zeta)(1+J (1-\zeta) )} $.   This provides the microscopic limit on $(\upsilon,\zeta)$.  The corresponding macroscopic limit is $2  \cV  \approx  1+2J  (1- \cZ) - 2  \sqrt{ J(1- \cZ)(1+ J(1- \cZ))} $ which can be solved to find
$J  \approx N_A (1-2 \cV)^2/(8 \cV \bdelZ)$ for $\cV<1/2 $, i.e. the minimum $J$ required to produce the observed $\cV$ and $\bdelZ$.

With the squeezed thermal state, $2S_0,S_x$ and $\var(S_z)$ play the role of $N_A, J_z$ and $\var(J_x)$, respectively, and we have $\bdelZ = \expect{S_0-S_x} =  (\ns + \nth + 2 \ns \nth)/2$ and $\cV = \var({S_z)}/S_0 = \frac{1}{2} \nc (1 + 2 \ns - 2 \sqrt{\ns(\ns+1)})(1 + 2 \nth)/(\nc + \ns + \nth + 2 \ns \nth)$.  In the macroscopic limit $\nc\rightarrow \infty$, we find $k  \propto 2\expect{S_0}$, and thus the size of the entangled groups is macroscopic. 
}

~

\renewcommand{\th}{T}
\PRLsection{correlation functions}
From \cite{ScullyZubairy1997} pp. 92-94, we can evaluate normally-ordered correlations in terms of the Wigner distribution $W$ as
\bea
\label{eq:Enm}
E_{m,n} &\equiv&  \expect{(a^\dagger)^m a^n}  = \int \!\! dx\,dp\, W(x,p) O_{m,n}(x,p), \\
O_{m,n} &\equiv& \lim_{\beta,\beta^*\rightarrow0} \left. \left[ {\partial_\beta} + \frac{\beta^*}{2}\right]^m  \left[ \partial_ {\beta^*} + \frac{\beta}{2}\right]^n \frac{e^{i \beta^* \alpha + i \beta \alpha^*}}{i^{m+n}}, \right. 
\eea
$\alpha \equiv x+ip$. This evaluates to the efficiently computed 
 $O_{m,n>m}= \alpha^{n-m} \sum_{a=0}^{m} |\alpha|^{2a}  \Upsilon_{m,n,a}$, $O_{m,n<m} = O_{n,m}^*$ where
\be \Upsilon_{m,n,a} \equiv  {(-{2})^{a-m} n! m! }/[{a! (a+n-m)! (m-a)!}]^{}.\ee  
Using the Wigner distribution of a squeezed thermal state
\be
\label{eq:Wigner}
W(x,p) = \frac{2 }{\pi\th^2} \exp[{-\frac{2 x^2}{\th^2 \asp^{2} }}] \exp[{-\frac{2 p^2 \asp^{2}}{\th^2 }}],
\ee
and recalling $\asp = \sqrt{1 + \ns} + \sqrt{\ns}$, $T^2 = 1+2 \nth$, we find
$
E_{m,n}  =   \sum_{a=0}^{m}   \Upsilon_{m,n,a}  I_{m,n,a}
$ 
where 
\bea
I_{m,n,a} &\equiv& \int dx dp \, W(x,p) (x+i p)^{n-m}  (x^2 + p^2)^a.
 \eea
\newcommand{\ksym}{t}
\newcommand{\lsym}{u}
Defining $\ksym \equiv (n-m)/2$, these evaluate to 
 \bea 
I_{m,n,a} &=&  \sum_{\lsym=0}^\ksym \sum_{b=0}^a \choose{2\ksym}{2\lsym}  \choose{a}{b} \frac{(-1)^{\ksym-\lsym}}{4^{\ksym+a} }
  \asp^{2(2\lsym+2b-\ksym-a )}   T^{2(\ksym+a)} \nnt (2(\ksym-\lsym+a-b)-1)!! (2(\lsym+b)-1)!!
 \eea
for $n-m$ even, and zero otherwise. 

~

{
\PRLsection{Reduced density matrix}
We calculate 
\be
R^{(2,N)}_{\ket{p_1,p_2},\bra{p'_1,p'_2}} =\sum_{\left\{p_{3} \ldots p_N \right\} 
\in \{H,V\}
 } \expect{a^\dagger_{p_1}a^\dagger_{p_2}  a^\dagger_{p_3} \ldots  a_{p_3}  a_{p_2'}a_{p_1'}}
 \ee
with respect to the state in Eq. (\ref{eq:RhoDef}). Defining $r,s$ as the number of $V$ polarizations in $\{p_1,p_2\}$ and $\{p'_1,p'_2\}$, respectively, we find
 \bea 
 R^{(2,N)}_{\ket{p_1,p_2},\bra{p'_1,p'_2}} &=& \sum_{m=0}^{N-2} \choose{N-2}{m} (\alpha_H^*)^{N-m-r} \alpha_H^{N-m-s} 
 \nnt
  E_{r+m,s+m} .
 \eea
}

\newcommand{\nsone}{\bar{n}_s}
\newcommand{\nstwo}{\bar{\bar{n}}_s}

\newcommand{\Perm}{P}


\end{document}